\documentclass[12pt,prb,superscriptaddress,showpacs,amsmath,amssymb, nofootinbib]{revtex4}
\usepackage{amsfonts}
\usepackage{bm}
\usepackage{verbatim}
\usepackage{color}

\usepackage{graphicx}
\newcommand{\im}{i}
\newcommand{\id}{1\!\!1}
\newcommand{\doo}{\partial}
\newcommand{\jaakko}[1]{\textcolor{black}{#1}}

\begin{document}

\title{Tetrads in solids: from elasticity theory to topological quantum Hall systems  and Weyl fermions}

\author{J.~Nissinen}
\affiliation{Low Temperature Laboratory, Aalto University,  P.O. Box 15100, FI-00076 Aalto, Finland}

\author{G.E.~Volovik}
\affiliation{Low Temperature Laboratory, Aalto University,  P.O. Box 15100, FI-00076 Aalto, Finland}
\affiliation{Landau Institute for Theoretical Physics, acad. Semyonov av., 1a, 142432,
Chernogolovka, Russia}

\date{\today}

\begin{abstract}
Theory of elasticity in topological insulators has many common features with relativistic quantum fields interacting with gravitational field in the tetrad form.  Here we discuss several issues in the effective topological (pseudo)electromagnetic response in three-dimensional weak crystalline topological insulators with no time-reversal symmetry that feature elasticity tetrads, including a mixed ``axial-gravitational" anomaly. This response has some resemblance to ``quasitopological" terms proposed for massless Weyl quasiparticles with separate, emergent fermion tetrads. As an example, we discuss the chiral/axial anomaly in superfluid $^3$He-A. 
 We demonstrate the principal difference between the elasticity tetrads and the Weyl  fermion tetrads in the construction of the topological terms in the action. In particular, the topological action expressed in terms of the elasticity tetrads, cannot be expressed in terms of the Weyl  fermion tetrads since in this case the gauge invariance is lost.

  \end{abstract}

\maketitle

\section{Introduction}

There are several sources of emergent tetrad gravity in solids. The tetrad field in particular emerges in Dirac and Weyl semimetals in the vicinity of nodes in the electron spectrum (the  fermion tetrads or Weyl tetrads). A different set of tetrads (the elasticity tetrads) emerge in the theory of elasticity, see e.g. Refs. \onlinecite{DzyalVol1980, LL7}. Our  consideration of the tetrad fields is based on approach formulated in two books by Landau and Lifshitz \cite{LL2,LL7} from their multi-volume ``Course of Theoretical Physics''. Lev Petrovich Pitaevskii was not only one who had completed the course but later also the editor of the books in the series.

Here we consider the  elasticity tetrad fields in weak topological insulators and  the fermion tetrads in related Weyl fermion systems. More specifically, we discuss the role of the crystalline tetrads in the response of 3+1-dimensional ($D=3+1$) weak topological insulators with anomalous quantum Hall effect (AQHE) and Weyl semimetals/superfluids with the chiral/axial anomaly. For the topological AQHE response, we obtain a $D = 2n+2=6$ mixed axial-gravitational anomaly, and consider its dimensional reduction to $D= 2n+1=5$ and $D = 2n=4$ anomalous actions,  as well as the extension to driven Floquet-Bloch systems, which are expressed in terms of three or four integer topological invariants in the crystal 4-momentum space.  We also note the possibility of  an emergent fermion tetrad gravity that is different for left- and right-handed Weyl fermions. This is possible in condensed matter systems although usually precluded by discrete symmetries. In the high-energy particle physics vacuum, the effect is constrained by (discrete) Lorentz symmetries.

 We do not consider the interplay of the effect of elastic deformations on the Weyl and Dirac fermions tetrads that has been considered e.g. in Refs. \onlinecite{Vozmediano2007, Vozmediano2012, VolovikZubkov2014,CortijoZubkov2016}.

\section{Weyl tetrads and chiral anomaly}

In this section we consider the Weyl fermion tetrads.

\subsection{Triads and tetrads in Weyl physics}

Let us start by a brief review of Weyl fermions in (quasi)relativistic theories. Take a massless, relativistic spin-1/2 particle with 4-momentum  $k^{\alpha}=(E,0,0,k^3)$. The Weyl equation is the statement
\begin{align}
(-E\sigma^0 \mp ck^3\sigma^3)\Psi_{\rm R,L}(k^3) = 0 =k_{\alpha}\sigma_{\rm R,L}^{\alpha} \Psi_{\rm R,L}(k^{\alpha})= 0, \label{eq:Weyl}
\end{align}
where $\sigma^{\alpha} = (1\!\!1, \sigma^a)$ denotes the $2\times2$ unit and Pauli matrices, $ k^{\alpha}k_{\alpha} \equiv  \eta_{\alpha \beta} k^{\alpha}k^{\beta} = 0$, and $c$ is the (effective) speed of light. There are two helicities $\sigma^3 = \pm 1$ for a massless particle $\Psi_{\rm R,L}(k^3) = (\Psi_{+\rm R,L}(k^3), \Psi_{- \rm R,L}(k^3))^{\rm T}$, corresponding to spin $S^3 = \sigma^3/2$ parallel or antiparallel to the momentum. The flavors $\rm R, L$ corresponding to the signs $\pm$ on the LHS of Eq. \eqref{eq:Weyl} label spin-1/2 particles that transform differently under a Lorentz boost $\Psi_{\rm R, L}(\vec{p})  = e^{\im \vec{K}_{\pm}\cdot \vec{\lambda}(p)}\Psi_{\rm R,L}(k^3)$, where $\im \vec{K}_{\pm} = \pm \vec{\sigma}/2$. In fact, the right and left fermions transform as complex conjugates  (up to a similarity transform). Boosting the LHS of Eq. \eqref{eq:Weyl}, one obtains the RHS of Eq. \eqref{eq:Weyl} in an arbitrary frame $k^\alpha = (E, k^1, k^2, k^3)$. 

Typically, \jaakko{even in particle physics}, the Weyl fermions are considered in the context of Dirac fermions, which are described by $\gamma$-matrices. The two-dimensional left and right handed Weyl fermions can be put into the 
4$\times$4 Dirac spinor representation, with $\Psi = \left( \Psi_{\rm L}, \Psi_{\rm R}\right)$ and
\begin{align}
\gamma^{\alpha} p_{\mu} \Psi = 0, \quad \{\gamma^{\alpha}, \gamma^{\beta}\} = 2 \eta^{\alpha\beta}, \quad  \gamma^{\alpha} = \left(\begin{matrix} & \sigma^{\alpha} \\ \overline{\sigma}^{\alpha} \end{matrix}\right), \quad \gamma^5 = \left( \begin{matrix} -\id & \\ & \id \end{matrix}\right), \label{eq:Weyl_basis}
\end{align} 
where $\overline{\sigma}^{\alpha} = (\id, -\sigma^a)$ and $\Psi_{\rm R, L} = \tfrac{\id\pm\gamma^5}{2} \Psi$. Under parity $P:\mathbf{x} \to -\mathbf{x}$, the left and right handed fermions interchange $P:\Psi \to \gamma^0 \Psi$. Similarly under time-reversal, $T: (t,\mathbf{x}) \to (-t, \mathbf{x})$ and $T: \Psi_{\rm R,L} \to \im\sigma^2 \Psi^{*}_{\rm R,L}$. Finally, for a charged particle we can define the charge conjugation $C: \Psi_{\rm R,L} \to \mp \im\sigma^2 \Psi_{L,R}^*$. These combine to give $\pm \gamma^5$ under the any combination of all three transformations $C,P,T$. Since by construction $\gamma^5 = \im \gamma^0 \gamma^1 \gamma^2 \gamma^3$ anticommutes with the $\gamma^\alpha$, $\alpha =0,1,2,3$, $CPT$ remains a symmetry of the Weyl system. Note also that either the anti-unitary symmetry $T^2 = (-1)^F$ or $(CT)^2 =(-1)^{F}$, where $F$ is the fermion number, guarantee the two-fold degeneracy of the spectrum at $E_{\rm R,L}=\pm c\vert \mathbf{p} \vert$. \jaakko{For Fermi superfluid and superconducting systems the charge conjugation is not well-defined but one has the anti-unitary particle-hole symmetry}. Recently fermionic systems with $(CT)^2 = (-1)^F$ and $T^2=1$, essentially interchanging the representations of the discrete symmetries, have been considered as well. This amounts to the ambiguity of two inequivalent anti-unitary time-reversal symmetries, $T$ and $CT \simeq (CTP) P^{-1}$, under emergent Lorentz invariance and gauge fields. 

In condensed matter, see e.g. reviews in Refs. \onlinecite{Burkov2018a} and \onlinecite{Mele2018}, the quasirelativistic equivalent of a Weyl Hamiltonian can arise in 3+1d fermionic systems with topologically protected nodes in the spectrum ---
 the monopoles in the Berry phase \cite{Volovik1987}. This can occur in topological superfluids, superconductors and semimetals \cite{NielsenNinomiya83, Volovik2003, Mele2018}. Depending on the physical symmetries and the (emergent) gauge fields at the node, the discrete symmetry transformations may take different forms, as noted above. Moreover, it is possible to consider discrete crystalline symmetries as well, leading to even more possibilities. The Hamiltonian $H(\mathbf{p})$ describing emergent Weyl fermions near the node in momentum space $\mathbf{p}$ contains effective triad field $e^k_{a}$, effective or real gauge field $A_k$, and dimensionless charge $q$:\cite{FrogNielBook,Volovik2003,Horava2005} 
\begin{align}
H(\mathbf{p}) =e^{k}_a\sigma^a(p_k-qA_k)
\,.
 \label{eq:HamiltonianWeyl}
\end{align}
Here, as above the $\sigma^a$ with $a=1,2,3$ are the standard Pauli matrices labeling spin, pseudospin, particle-hole or band index depending on the context.
The four dimensional tetrad field $e^{\mu}_{\alpha}$ is relevant when the fermion Green's function, or the action is considered:
\begin{align}
G^{-1}(p_{\mu}) =e^{\mu}_\alpha\sigma^\alpha(p_\mu-qA_\mu)
\,,
 \label{eq:GreenWeyl}
\end{align}
where $p_{\mu} = (\omega, -\mathbf{p})$, $\alpha=0,1,2,3$ and $\sigma^{\alpha} =(\id, \sigma^a)$  as above. We call  the fermion tetrad $e^{\mu}_{\alpha}$ the Weyl tetrad --- to distinguish them from the elasticity tetrads discussed later.

In Eq.(\ref{eq:GreenWeyl}), $e^0_{\alpha}$ is dimensionless, whereas $e^a_{k}$ has dimensions of velocity (\jaakko{or dimensionless in units where $c=1$}).  If there is no special symmetry between the Weyl nodes in the spectrum, the tetrads for the left and right handed fermions can have in principle different tetrads. And of course, the left and right tetrads have opposite signs of determinant, $e \equiv \det e^{\alpha}_{\mu}$. The topological Lifshitz transition with the change of the sign of the determinant --- the transition to antispacetime via the space time with the degenerate tetrads --- is described in Ref.\onlinecite{NissinenVolovik2018}.

\subsection{Gravity and  chiral Weyl fermions: which one is more fundamental?}

In the Standard Model of particle physics, massive 4-component Dirac fermions with $SU(2)_{\rm L} \times SU(2)_R$ spin $(1/2, 1/2) = (1/2,0)_{\rm L} \oplus (0,1/2)_{\rm R}$ represent a $PT$-invariant mixture of chiral, massless left handed (isospin doublet) and right handed (isospin singlet) Weyl fermions due to the electroweak Higgs mechanism. \jaakko{In Nature}, the chiral fermions are primary objects, while the Dirac fermions are secondary \jaakko{composite} objects. \jaakko{In contrast, the condensed matter} chiral fermions are not elementary objects but instead emerge together with effective gravity and effective gauge fields in the vicinity of the momentum space Berry phase monopole at the node.

From the condensed matter point of view, the chiral fermions of the massless Standard Model  with their tetrad (semiclassical) gravity are not necessarily the primary  objects, and the question  
"What is more fundamental: gravity or the chiral Weyl fermions?" becomes appropriate \cite{ChangSoo96}.

Two scenarios are possible depending on what is primary:
\begin{itemize}
\item[(i)] If gravity is primary, then there should be the unique tetrad field $e^{\mu}_{\alpha}$, which is the same for both chiralities.
This means that the two $SU(2)_{\rm L,R}$ Lorentz representations are complex conjugate, $\im K^a_{\rm L,R} = \pm \sigma_{\rm L,R}^a/2$ with $a=1,2,3$, and must have different signs for right and left handed fermions.

\item[(ii)] If the chiral left and right handed fermions are primary, they may have different tetrad fields, $e^{\mu}_{\alpha \rm R} \neq e^{\mu}_{\alpha \rm L} $. If gravity has a unique spacetime geometry, there is a constraint that  the two tetrads should comprise the same metric field:
\begin{align}
g^{\mu\nu}=e^{\mu}_{a{\rm R}}e^{\nu}_{b{\rm R}}\eta^{ab}=e^{\mu}_{a{\rm L}}e^{\nu}_{b{\rm L}}\eta^{ab}\,,
 \label{eq:LeftRightSymmtetry}
\end{align}
which is invariant under the local Lorentz transformations $e^{\mu}_{\alpha\rm R,L} \to \Lambda^{\beta}_{\alpha} e^{\mu}_{\beta\rm R,L}$. In order to form a massive Dirac fermion, the tetrads must be connected by the discrete $P, T, C$ symmetries that transform the left and right handed fermions into each other. For example, under parity
$e^{\mu}_{0{\rm R}}=e^{\mu}_{0{\rm L}}$, $e^{\mu}_{1{\rm R}}=-e^{\mu}_{1{\rm L}}$, etc. 
\end{itemize}

Let us consider how these two scenarios lead to Dirac fermions in more detail.

In the case (i), the Lagrangians for right and left Weyl  fermions are expressed in terms of the same tetrads. Taking for simplicity only the diagonal tetrad matrices $e^{\mu}_{\alpha} = (e^0_0, e^k_a)$ one has:
\begin{align}
\mathcal{L}_{\rm R}=\Psi^+_{\rm R}(e^{0}_{0}p_0 + e^{k}_a\sigma^a p_k)\Psi_{\rm R}\,\,,\,\, 
\mathcal{L}_{\rm L}=\Psi^+_{\rm L}(e^{0}_{0}p_0 - e^{k}_a\sigma^a p_k)\Psi_{\rm L}
\,.
 \label{eq:Weyl1}
\end{align}
The massive Dirac fermions are obtained  after the electroweak Higgs transition when, due to the broken electroweak symmetry, matrix elements  $M$ between the chiral right and left handed fermions appear.  The Lagrangian for Dirac fermions is
\begin{eqnarray}
\mathcal{L}_{\rm Dirac}=\Psi^+
\left( \begin{array}{cc}
e^{0}_{0}p_0 + e^{k}_a\sigma^a p_k &M\\
M&e^{0}_{0}p_0 -  e^{k}_a\sigma^a p_k 
\end{array} \right)\Psi =
\label{eq:Dirac1}
\\
=\Psi^+\left(e^{0}_{0}p_0 + \tau_3 e^{k}_a\sigma^a p_k + M\tau_1\right)\Psi =
\label{eq:DiracNoncomplete}
\\
=\bar\Psi (\gamma^\alpha e^\mu_{\alpha} p_\mu +M)\Psi
\,,
\label{eq:Dirac}
\end{eqnarray}
where the Pauli matrices $\tau^a$ operate in the $\rm L,R$-chirality space and the Dirac matrices $\{\gamma^\alpha, \gamma^{\beta}\} = 2\eta^{\alpha\beta}$ with
\begin{equation}
\bar\Psi =\Psi^+\gamma^0 \,\,,
\,\, \gamma^0=\tau_1
 \,\,,   \,\,  \gamma^a=i\tau_2 \sigma^a\,\,,\,\, a=1,2,3
\,.
\label{gamma}
\end{equation}
\jaakko{This is nothing else than the chiral Weyl gamma matrix basis in Eq. \eqref{eq:Weyl_basis}}.

In the case (ii), the chiral fermions are fundamental, and they generate tetrad fields, which are different for the two chiralities:
\begin{align}
\mathcal{L}'_{\rm R}=\Psi^+_{\rm R}(e^{0{\rm R}}_{0}p_0 + e^{k{\rm R}}_a\sigma^a p_k)\Psi_{\rm R}\,\,,\,\, 
\mathcal{L}'_{\rm L}=\Psi^+_{\rm L}(e^{0{\rm L}}_{0}p_0 + e^{k{\rm L}}_a\sigma^a p_k)\Psi_{\rm L} \,,
 \label{eq:Weyl2}
\end{align}
where $e^{\mu}_{\alpha \rm R} \neq e^{\mu}_{\alpha \rm L}$. The parity symmetric Dirac Lagrangian in Eq.(\ref{eq:Dirac}) is obtained if there is an underlying symmetry between the right and left fermions, which leads to $e^{0}_{0{\rm R}}=e^{0}_{0{\rm L}}\equiv e^{0}_{0}$ and 
$e^{k}_{a{\rm R}}=-e^{k}_{a{\rm L}} \equiv e^{k}_{a}$ for $a=1,2,3$. Then one has
\begin{eqnarray}
\mathcal{L}_{\rm Dirac}=\Psi^+
\left( \begin{array}{cc}
e^{{\rm R}0}_{0}p_0 + e^{{\rm R}k}_a\sigma^a p_k &M\\
M&e^{{\rm L}0}_{0}p_0 +  e^{{\rm L}k}_a\sigma^a p_k 
\end{array} \right)\Psi 
\\
=\Psi^+\left(e^{0}_{0}p_0 + \tau_3 e^{k}_a\sigma^a p_k + M\tau_1\right)\Psi=\bar\Psi (\gamma^a e^\mu_a p_\mu +M)\Psi
\,,
\label{eq:Dirac2}
\end{eqnarray}
i.e. the same Dirac equation (\ref{eq:Dirac}).

\subsection{Chiral anomaly}

The chiral anomaly in momentum space has been observed in $^3$He-A\cite{Bevan1997} and discussed in Weyl semimetals, \jaakko{ see e.g. the original paper by Nielsen and Ninomiya \cite{NielsenNinomiya83} and Refs. \onlinecite{Burkov2018a, Mele2018}}. The anomaly is the non-conservation of the chiral/axial current 
\begin{align}
J^5_{\mu} = \overline{\Psi} \gamma^5 \gamma_{\mu} \Psi = \overline{\Psi}_R \gamma_\mu \Psi_R - \overline{\Psi}_L \gamma_\mu \Psi_L
\end{align}
due to a  real or effective gauge field $A_{\mu}$ for massless Weyl fermions,
\begin{align}
\partial_{\mu} J^{5\mu} =   \frac{1}{16 \pi^2}q^2\epsilon^{\mu\nu \alpha \beta} F_{\mu\nu}F_{\alpha\beta}, \quad J^{5\mu} = \frac{1}{4 \pi^2}q^2\epsilon^{\mu
\nu \alpha \beta} A_{\nu}\partial_{\alpha} A_{\beta} \,. \label{eq:chiral_current}
\end{align}
In the neutral superfluid $^3$He-A, the effective vector potential is $A_\mu({\bf r},t) =(0, p_F \hat{\mathbf{l}}({\bf r},t))$ with $q=\pm 1$ is the effective axial charge of the Weyl quasiparticles. In $^3$He-A,  $A_{\mu}$ is a dynamical field featuring the orbital degrees of freedom of $p$-wave the superfluid condensate, and  the quantity $\pm p_F \hat{\mathbf{l}}({\bf r},t)$ is the shift of the position of the Weyl node in the orbital texture \cite{Volovik2003}.
In QED, $A_\mu$ is the electromagnetic field of the local $U(1)$ symmetry,
and $q=-1$ for the electric charge of electron.
In Weyl semimetals with electronic quasiparticles, both electromagnetic and effective axial gauge fields contribute to the anomaly equation: the axial gauge field coupling to the Weyl fermions is the shift of the Weyl nodes $q_{\mu}({\bf r},t)$ in 4-momentum space, caused by \jaakko{external perturbations e.g. spin-orbit interaction, exchange fields, strain etc. in the material}.

In $^3$He-A and Weyl semimetals the chiral current $J_{\mu}^5$ is not a physical quantity, since the left and right quasiparticles are not well defined quantities in the full theory: the different chiralities are determined only at low energy in the vicinity of the Weyl nodes. 
 
Nevertheless, the creation of the left and and right quasiparticles by the chiral anomaly is accompanied by the creation of the linear momentum from the the ground state of the system by the field strength $F_{\mu\nu} \neq 0$ in Eq.\eqref{eq:chiral_current}. Since the left and right quasiparticles have different momenta (in $^3$He-A ${\bf p}_\pm = \pm p_F \hat{\bf l}$), there is a net production of linear momentum per unit time in an external field $F_{\mu\nu}$. The latter is an observable physical force. In $^3$He-A, this provides an extra force acting on skyrmions in the $\hat{\mathbf{l}}$-texture (the so-called Kopnin force) and was measured in Ref.\onlinecite{Bevan1997}. Moreover, the extra piece in the effective action of the Weyl superfluid that produces the equation of motion \eqref{eq:chiral_current} can be written as
\begin{align}
S_{\rm WZN} = \frac{1}{48\pi^2 \hbar^2}\int_{X^5} d \tau d^4 x~ \epsilon^{\kappa\mu\nu\lambda\rho} A_{\kappa}\partial_{\mu}A_{\nu}\partial_{\lambda} A_{\rho}.
\end{align}
where $X^5:(\tau, x^{\mu})$ is a fictitious five-dimensional manifold with extra coordinate $\tau$ with the property that $\doo X = X^4$ is the physical spacetime. The action $S_{\rm WZN}$ is a five-dimensional Chern-Simons action for the effective gauge field $A_{\mu}$; its variation is a pure boundary term in physical space, which produces the anomaly Eq. \eqref{eq:chiral_current}. 
The extension to five-dimensions implied by $S_{\rm WZN}$ is unambigous due to the integer quantization of the normal Fermi system $N_0 = \tfrac{4\pi}{3}p_F^3V$ is the total particle number, an integer \cite{Volovik1986} \jaakko{(see the Appendix for the topological quantization of the 3D CS action)}.

\section{Elasticity tetrads in anomalous Hall Insulators}

In this section we consider elasticity tetrads which emerge in the elasticity theory,  as applied in the the anomalous quantum Hall response of weak crystalline topological insulators.

\subsection{Chern-Simons action and topological invariants}

\jaakko{Consider a band insulator with broken time-reversal invariance in three dimensions. There are two possibilities: a trivial insulator and a weak topological insulator with an anomalous Hall effect, protected by crystalline translational symmetries, see e.g. the recent review in Ref. \onlinecite{QAHall2018}. That is, the electromagnetic response of the $D=3+1$-dimensional insulator may contain the following topological term:}
\begin{equation}
S_{4D}[A_{\mu}]=\frac{1}{4\pi^2} \sum_{a=1}^3N_a  \int d^4 x~ E^{~a}_{\mu} \epsilon^{\mu\nu\alpha\beta} A_\nu \partial_\alpha A_\beta\,.
\label{action}
\end{equation}
Here the $E_{\mu}^{~a}(x) = (E_0^{~a}(x), E^{~a}_i(x) )$, with $i,a=1,2,3$ play the role of tetrad field with dimensions of momentum $[L]^{-1}$. The spatial $\mathbf{E}^a(x) = E^{~a}_{i}(x)$ are primitive  vectors of the  reciprocal Bravais lattice, which depend on spacetime  coordinates under deformations of the crystal lattice. 
As in the case of the chiral anomaly, the prefactors in the topological response are expressed in terms of the
integer topological momentum space invariants \cite{KaplanEtAl93}. In our case we have the integer coefficients $N_a$ -- three momentum space invariants, i.e. winding numbers, expressed in terms of integrals of the Green's functions:
\begin{equation}
N_a=\frac{1}{8\pi^2}\epsilon_{ijk} \int_{-\infty}^{\infty} d\omega\int_{\rm BZ} dS_a^i~{\rm Tr} [(G\doo_{\omega} G^{-1}) (G\doo_{k_i} G^{-1}) ( G\doo_{k_j} G^{-1})]\,.
\label{Invariants}
\end{equation}
The momentum integral is over the 2D torus in the cross section $\mathbf{S}_a$ of the three-dimensional Brillouin zone (BZ). The integer invariants $N_a$ are topological  invariants of the system and in particular remain locally well-defined under smooth deformations of the lattice. Under sufficiently strong deformations or disorder one can have regions of different $N_a(x)$ with chiral edge modes. In that case, the global invariant, if any, is defined by the topological charge of the dominating  cluster which percolates through the system \cite{Volovik2018}.

\subsection{Elastic deformations and gauge invariance}

The matrix $E^{~a}_\mu(x)$ can be taken as the hydrodynamic variable of elasticity theory, which plays the role of the tetrad (with dimensions $[L]^{-1}$) in effective theory \cite{DzyalVol1980}. We call them the elasticity tetrads. For deformed lattices, it depends on space and time,  $E^{~a}_\mu(x^{\nu})$, $x^\nu=({\bf r},t) = (x^i,t)$. Non-constant, non-quantized parameters in the Chern-Simons action violate gauge invariance. But in topological insulators this does not happen, under deformations Eq. (\ref{action}) remains gauge invariant even for space and time dependent  tetrad parameters $E^{~a}_\mu(x)$. 

The reason is the following. Under deformation of the crystal lattice, but in the absence of dislocations, $E^{~a}_\mu(x)$ is an exact differential. In detail, it can be expressed in general form in terms of a system of three deformed crystallographic coordinate planes, surfaces of constant phase $X^a(x)=2\pi n^a$, $n^a \in \mathbb{Z}$ with $a=1,2,3$. The intersection of three constant surfaces
\begin{equation}
X^1({\bf r},t)=2\pi n^1 \,\,, \,\,  X^2({\bf r},t)=2\pi n^2 \,\,, \,\, X^3({\bf r},t)=2\pi n^3 \,,
\label{points}
\end{equation}
\jaakko{are points of the (possibly deformed) crystal lattice, $L = \{ \mathbf{R} = {\bf r}(n_1,n_2,n_3) \vert n^a \in \mathbb{Z}^3\}$. For crystals, there is no periodicity in time $X^0(\mathbf{r},t) \equiv 0$ and in the undeformed case, $X^a(\mathbf{r},t) = \mathbf{K}^a \cdot \mathbf{r}$, where $\mathbf{K}^a \equiv E_i^{(0)a}$ are the (primitive) reciprocal lattice vectors spanning the BZ. The elasticity tetrads appearing in Eq. \eqref{action}  with units of crystal momentum are gradients of the phase function}
\begin{equation}
E^{~a}_\mu(x)= \partial_\mu X^a(x)\,,
\label{reciprocal}
\end{equation}
and in the absence of dislocations, the tensor $E^{~a}_\mu(x)$ satisfies the integrability condition:
\begin{equation}
\partial_\mu E^{~a}_\nu(x)-\partial_\nu E^{~a}_\mu(x)=0\,.
\label{integrability}
\end{equation}
This condition supports the gauge invariance of the action (\ref{action}) even in the presence of deformations:  the variation $\delta_{\phi} S$ of Eq. \eqref{action} under $\delta A_{\mu} = \partial_{\mu} \phi$ is identically zero in the presence of the constraint Eq. \eqref{integrability} (if the QH currents on the boundaries of the sample are not taken into consideration). 

That is why the inverse transformation \jaakko{$\mathbf{r}(X)$ of Eq. \eqref{points}} determine 
 three  primitive lattice vectors $E_{a}^{~m}(x)$, which depend on coordinates in the deformed state:
\begin{equation}
\jaakko{\frac{\partial r^m}{\partial X^a}=E_{~a}^{m}(x)}\,.
\label{InverseTransformation}
\end{equation}
They correspond to the inverse tetrads, since
\begin{equation}
E^{~a}_lE_{~a}^{m} =\delta^{m}_l\,.
\label{InverseTetrad}
\end{equation}
The determinant of the inverse tetrad is equal to the volume \jaakko{$V_{\rm cell}$ of the primitive unit cell} of the crystal:
\begin{equation}
{\rm det} E_a^m = V_{\textrm{cell}} \,\,\,, \,\, {\rm det} E^a_m  = V_{\rm BZ}= 
\frac{1}{V_{\textrm{cell}}} \,.
\label{DeterminantTetrad}
\end{equation}

\subsection{Invariants in terms of 4D integral in momentum space}
\label{Zubkov}

The three invariants can be written in the oovariant form suggested by Zubkov, see Ref. \cite{KhaidukovZubkov2017}:
\begin{equation}
S=\frac{1}{4\pi^2}   \int d^4 x~  \epsilon^{\mu\nu\alpha\beta} M_{\mu} A_\nu \partial_\alpha A_\beta\,,
\label{actionZubkov}
\end{equation}
where
\begin{equation}
M_{\mu}=\frac{i}{48\pi^2}\epsilon_{\mu\nu\alpha\beta} \int d^4p \,{\rm Tr}[(G\partial_{p_\nu} G^{-1}) (G\partial_{p_\alpha} G^{-1}) ( G\partial_{p_\beta} G^{-1})]\,.
\label{InvariantsZubkov}
\end{equation}
This was originally applied to Wilson fermions on the 4D lattice, where the $d^4p$ integral is over 4D torus. In our case $p_4=\omega$, with $-\infty <\omega<\infty$, and thus the component $M_4$ is not determined, since the integral over $p_4$ diverges. So this representation makes sense only for $\mu=1,2,3$, and we have
\begin{equation}
M_l= \sum_{a=1}^3 N_a E^{~a}_l\,.
\label{InvariantsZubkov2}
\end{equation}

\subsection{Explicit gauge invariant form}

After a partial integration, the action in Eq. (\ref{action}) takes the explicit gauge invariant form
\begin{align}
\frac{1}{16\pi^2} \sum_{a=1}^3 N_a \int_{M\times \mathbb{R}} d^4 x X^a \epsilon^{\mu\nu \alpha \beta} F_{\mu\nu} F_{\alpha \beta} \,,
\label{InvariantAction}
\end{align}
where the phase functions $X^a(x)$ play the role of three ``axion" fields.  In the non-distorted lattice one has $X^a(x)= \mathbf{K}^a \cdot \mathbf{r} = E^{(0)a}_i x^ i$, which leads to the linear divergence of integral in Eq.(\ref{InvariantAction}). This demonstrates that  even in the absence of deformation such description is not useful. \jaakko{The periodicity of the lattice in Eq. \eqref{points} in space can be used to restrict this divergence to a finite three torus $T^3$ corresponding to the lattice with a fixed origin}. The invariance of the action under the periodicity $X^a \to X^a + 2\pi n^a$ of the Bravais lattice requires an integer value of the integral
\begin{align}
\frac{1}{8 \pi} \sum_a  N_a n^a\int_{T^3\times \mathbb{R}} d^4 x \, \epsilon^{\mu\nu \alpha \beta} F_{\mu\nu} F_{\alpha \beta} \,,
\label{FluxQuantization}
\end{align}
 which is quantized for $U(1)$ gauge fields, see the Appendix. However, such requirement is artificial for the (background) gauge field $F_{\mu\nu}$ which does not need to respect the periodicity of the lattice $T^3$ but is rather a general gauge field with no periodicity  (in contrast e.g. to Berry curvature). That is why the action (\ref{action}) is more appropriate.

\subsection{Hall conductivity}

The Chern-Simons action Eq. \eqref{action} gives the conductivity tensor
\cite{Halperin1987}
\begin{equation}
\sigma_{ij}=\frac{e^2}{2\pi \hbar}\epsilon_{ijk}G_k\,,
\label{conductivity}
\end{equation}
where $G_k$ is a reciprocal lattice vector. It is expressed in terms of the topological invariants $N_a$:
\begin{equation}
G_k=\sum_a N_a E^a_k\,,
\label{ReciprocalVector}
\end{equation}
where $E^{~a}_{k}$ is a \jaakko{primitive} reciprocal lattice vector. The Hall conductivity is dissipationless and fully reversible.

\subsection{Chiral magnetic effect}

The action (\ref{action}) also describes the chiral magnetic effect (CME) \cite{FukushimaEtAl08, LiEtAl16}, i.e. an electric current along an applied magnetic field:
\begin{equation}
{\bf J}=\frac{ {\bf B}}{2\pi^2}\sum_{a=1}^3 N_aE^a_4 \,.
\label{conductivity}
\end{equation}
This current contains $E^{~a}_4=\partial X^a/\partial t$ and thus it vanishes in equilibrium in agreement with Bloch theorem, according to which in the ground state of the system or in the equilibrium state in general, the total current is absent, see e.g. Ref. \onlinecite{Yamamoto2015}.

\subsection{Bragg glasses and quasicrystals}

 The action in Eq.  (\ref{action}) can be also applied to Bragg glasses, the disordered state without dislocations \cite{Giamarchi1995}, since the  constraint Eq. (\ref{integrability}) is obeyed.   The \jaakko{non-topological point defects in crystals (impurities and vacancies) do not destroy the topological response}. One may apply Eq. (\ref{action}) to topological quasicrystals\cite{TopQuasi2016} \jaakko{obtained by dimensional reduction from a regular six-dimensional space crystal. In that case $a$ ranges from 1 to 6}. 

\subsection{Effective gravity}

The physical meaning of the constraint Eq. (\ref{integrability}) becomes clear when the action Eq. (\ref{action}) is expressed in terms of gravity and gauge fields. The components $E^{~a}_i$ play the role of the triad field \jaakko{(with the units of reciprocal space)} in the geometric description of deformations in elasticity theory, see e.g. Ref. \onlinecite{DzyalVol1980}. \jaakko{The effective metric determines the real distance $d l$ after a deformation $R^m(x)$ between the atoms at positions $x^i$, $x^i+dx^i$ in the undeformed state}:
\begin{equation}
dl^2=g_{ik}dx^i dx^k\,,
\label{distance}
\end{equation}
where
\begin{equation}
g_{ik}= \frac{dR^m}{dx^i}\frac{dR^m}{dx^k}     \,.
\label{metric1}
\end{equation}
Under \jaakko{(a continuous)} deformation, the tetrad of the non-deformed state $E^{(0)a}_i$ transforms to the general tetrad 
$E^{~a}_{i}$ according to coordinate transformation:
\begin{equation}
E^{~a}_i = \frac{d R^k}{dx^i} E^{(0)a}_k    \,.
\label{Transformation}
\end{equation}
\jaakko{The vector index $a$ corresponds to basis of lattice vectors}. This gives
\begin{equation}
g_{ik}= \frac{dR^m}{dx^i}\frac{dR^m}{dx^k}
= \eta_{ab} E^a_i E^b_k     \,.
\label{metric2}
\end{equation}
with  $\eta_{ab}$ expressed in terms of the primitive vectors of the \jaakko{undeformed} crystal:
\begin{equation}
\eta_{ab} = E_a^{(0)i} E_b^{(0)i}   \,\, \,.
\label{eta}
\end{equation}
The dimension of $\eta_{ab}$ is the square of length, $[\eta_{ab}]=[L]^2$. \jaakko{Note that $\eta_{ab}$ is not necessarily the Lorentz metric, ${\rm diag}(1,-1,-1,-1)$ with $c=1$, but can be arbitrary symmetric matrix corresponding to the metric of the undeformed state \cite{LL2}}.  Moreover, $\eta_{ab}$ is not dimensionless but carries the dimensions of length. \jaakko{Similarly, the tetrad $E^{~a}_i$ differs from the standard dimensionless tetrad by a primitive reciprocal lattice vector $E^{(0)a}_i$ in Eq. \eqref{Transformation}}.

\subsection{Dislocations and torsion}

In the presence of the  topological defects corresponding to dislocations, the density of dislocations plays the role of torsion:
\begin{equation}
T^a_{kl} =(\partial_k E^{~a}_l-\partial_l E^{~a}_k) \,\,,\,\,  T^m_{kl}=E_a^{~m}(\partial_k E^{~a}_l-\partial_l E^{~a}_k)\,.
\label{Torsion}
\end{equation}
\jaakko{Note that this differs from the differential geometric torsion by a factor corresponding to the primitive lattice vectors $E^{(0)a}_i$}. For a single dislocation one obtains, as in the case of a single $2\pi$ vortex \jaakko{(see e.g. Ref. \onlinecite{BeekmanEtAl17} and references therein)},
 \begin{equation}
\partial_k E^{~a}_l-\partial_l E^{~a}_k=n^a \epsilon_{klp}\int ds \,\kappa^{p} \delta({\bf r}-{\bf r}(s,t))\,.
\label{dislocation}
\end{equation}
Here the $n^a$ are \jaakko{integer} topological charges of the dislocation --- ``winding numbers", which enter the  Burgers vector of dislocation, $b^m=\sum_a E_a^{~m} n^a$, and $\kappa^{p}$ is the unit tangent vector along the corresponding dislocation line:
\begin{equation}
{\mbox{\boldmath$\kappa$}}=d{\bf r}/ds\,.
\label{kappa}
\end{equation}
The torsion is expressed in terms of the  Burgers vector of dislocation:
\begin{equation}
T^m_{kl}= E_{a}^{~m} n^a \epsilon_{klp}\int ds \,\kappa^{p} \delta({\bf r}-{\bf r}(s,t)) = b^m \epsilon_{klp}\int ds \,\kappa^{p} \delta({\bf r}-{\bf r}(s,t))  \,.
\label{DislocationTorsion}
\end{equation}

\subsection{Dislocation fermion zero modes and Callan-Harvey anomaly cancellation}

The action in Eq. \eqref{action} with the constraint Eq. (\ref{integrability}) represents a kind of mixed gravitational anomaly for gravity with zero curvature and zero torsion, see Sec.\ref{Sec.MixedAnomaly}.

The constraint Eq. \eqref{integrability} is violated in the presence of topological defects of the crystal lattice  such  as dislocations, stacking faults and twin boundaries. The fermion zero modes on these topological defects provide the Callan-Harvey mechanism of anomaly cancellation 
\cite{CallanHarvey85,RanEtAl09,Zaanen2012,Zaanen2014}.
The number of the 1D fermion zero modes on dislocation is the sum of the products of topological charges of the insulator $N_a$ and  topological charges (winding numbers) $n^a$ of the dislocation:
\begin{equation}
\nu_{\rm ZM} = \sum_a N_a n^a  \,.
\label{ZeroModes}
\end{equation}
 Note that this integer number is purely topological and does not contain the \jaakko{elasticity tetrads}.

 \subsection{Mixed anomaly}
 \label{Sec.MixedAnomaly}
 
The mixed anomaly corresponding to Eq. \eqref{action} in $D=4$ can be related to a topological term in $D=6$ dimensional spacetime in terms of the torsion in Eq. \eqref{Torsion} and $U(1)$ gauge field:
 \begin{align}
S_{6D}=\frac{1}{96\pi^2} \sum_{a=1}^n N_a \int_{X^6} d^6 x \,   \epsilon^{\mu\nu \alpha \beta\gamma\delta} 
T^a_{\gamma\delta}F_{\mu\nu} F_{\alpha \beta} \,.
\label{MixedAnomaly6D}
\end{align}
This anomaly is mixed, since it contains the real or effective gauge field acting on Weyl fermions, and the tetrads of elasticity gravity.

In Eq.(\ref{MixedAnomaly6D})  the number $n$ of the tetrad components $E^{~a}_{\gamma}$ describing the crystal ``planes" can be smaller than the dimension of space (if in the other,  missing directions the system is not periodic). This happens for example for smectic liquid crystals, where there is one set of planes $n=1$, and for vortex lattices, \cite{VolovikDotsenko1979}
with two sets of planes, $n=2$.
\jaakko{Here $n=3$ for the 3+1d topological insulators, while for e.g. Wilson fermions in Sec. \ref{Zubkov} and in Floquet crystals in Section \ref{Sec:FloquetCrystal}, one has $n=4$}.

Dimensional reduction of Eq.(\ref{MixedAnomaly6D}) to $D=5$ gives the \jaakko{mixed} Wess-Zumino term:
 \begin{align}
S_{5D}=\frac{1}{8\pi^2} \sum_{a=1}^n N_a \int_{X^5} d^5 x \,   \epsilon^{\mu\nu \alpha \beta\gamma} 
E^{~a}_{\gamma}F_{\mu\nu} F_{\alpha \beta} \,.
\label{MixedAnomaly5D}
\end{align}
In the absence of dislocations, the 5-form in the \jaakko{intergrand of} Eq. \eqref{MixedAnomaly5D} is not only a closed but an exact form, and transforms to a surface integral, i.e. to  Eq. \eqref{action}. This describes the dimensional reduction of Eq. \eqref{MixedAnomaly6D} to the $D=4$ action in Eq. \eqref{action} via the $D=5$ Wess-Zumino term in Eq. \eqref{MixedAnomaly5D}.

\subsection{Floquet crystal, CME and Chern-Simons action in tangent space}
\label{Sec:FloquetCrystal}

If one adds periodicity in time \jaakko{corresponding e.g. to a ``Floquet" drive}, an additional topological invariant  $N_4$ enters:
\begin{equation}
S_{\rm F}=\frac{1}{4\pi^2} \sum_{a=1}^4N_a  \int d^4 x~ E^{~a}_{\mu} \epsilon^{\mu\nu\alpha\beta} A_\nu \partial_\alpha A_\beta\,.
\label{action4d}
\end{equation}
In case of \jaakko{a Floquet-Bloch} crystal with a fixed drive frequency $\omega_{\rm F}$ one has 
$E^{~4}_{\mu}=\omega_{\rm F} \delta^{~4}_{\mu}$.
The invariant  $N_4$ breaks parity, and gives rise to the chiral magnetic effect (CME) \cite{FukushimaEtAl08,LiEtAl16}, i.e. to a current along an applied magnetic field:
\begin{equation}
{\bf J}=\frac{1}{2\pi^2}  N_4 \omega_{\rm F} {\bf B}\,.
\label{CME}
\end{equation}
This current does not violate the Bloch theorem, since it is proportional to frequency and disappears in the absence of the drive. This is in contrast to a similar term with a chiral chemical potential imbalance\cite{FukushimaEtAl08} that violates the Bloch theorem.

In the 4D periodic Floquet case one can introduce the gauge field $A_a$ in the  crystal frame with $a=0,1,2,3$ (see e.g. Ref. \onlinecite{Moritscht1995})  as
\begin{equation}
A_\mu = E^{~a}_\mu A_a \,,
\label{ADual}
\end{equation}
and if $E_{\mu}^{~a}$ satisfies the integrability constraint Eq. \eqref{integrability} (which is valid for the elasticity tetrads)
\begin{align}
F_{\mu\nu} =\partial_\nu ( E^{~a}_\mu  A_a)- \partial_\mu ( E^{~a}_\nu  A_a)=
E^{~a}_\mu\partial_\nu   A_a-  E^{~a}_\nu\partial_\mu   A_a \nonumber \\
= E^{~a}_\mu E^{~b}_\nu\partial_b   A_a-E^{~a}_\nu E^{~b}_\mu\partial_b   A_a
=E^{~a}_\mu E^{~b}_\nu F_{ab}
\,,
\label{FDual}
\end{align}
one gets
\begin{align}
S &=
\frac{1}{8\pi^2} N_a \epsilon^{\mu\nu\alpha\beta} \int d^4 x\,  E^{~a}_{\mu} E^{~b}_{\nu}   E^{~c}_{\alpha}   E^{~d}_{\beta}  A_b F_{cd}
\nonumber
\\
&=
\frac{1}{8\pi^2} N_a \epsilon^{abcd} \int d^4 x\,  {\rm det}(E)  A_b F_{cd}=
\frac{1}{8\pi^2} N_a \epsilon^{abcd} \int d^4 X\,  A_b F_{cd}
\,,
\label{actionDual}
\end{align}
The Eq.(\ref{actionDual}) is invariant under $A_a(X) \rightarrow A_a(X) + \partial \phi/\partial X_a$ \jaakko{since the four torus $T^4$ is a closed manifold}.

The invariants of Floquet crystals can be described also in terms of Eqs. (\ref{actionZubkov})
 and (\ref{InvariantsZubkov}), with all four \jaakko{non-zero} components of $M_\mu$:
\begin{equation}
M_\mu= \sum_{a=1}^4 N_a E^{~a}_\mu\,.
\label{InvariantsZubkov3}
\end{equation}

\subsection{Why the Chern-Simons \jaakko{AQHE} action does not exist in Weyl semimetal}

\jaakko{For Weyl semimetals the corresponding action Eq. \eqref{action} with QAHE can be found e.g. in Eq.(1) of Ref. \onlinecite{Burkov2018} and for relativistic Weyl fermions in Ref. \onlinecite{KlinkhamerVolovik05}. However, in semimetals such an action is not appropriate in general due to the violation of gauge invariance}.
In semimetals, the separation $q_\mu$ between the Weyl nodes \jaakko{can} depend on space and time under deformations, and one obtains:
\begin{equation}
S_{4D}=\frac{1}{4\pi^2} \int d^4 x~ q_\mu(x) \epsilon^{\mu\nu\alpha\beta} A_\nu \partial_\alpha A_\beta\,.
\label{actionWrong}
\end{equation}
The Eq. \eqref{actionWrong} violates gauge invariance, unless $q_\mu$ is a constant
of \jaakko{the topological medium/vacuum} or there is a integrability constraint $\doo_\nu q_\mu-\doo_\mu q_\nu=0$.\cite{Carroll1990}

Neither condition is true in general if $q_\mu$  is the 4-momentum space separation of the Weyl points. This shows that the description in terms of the action is not appropriate for the discussion of the 3D AQHE in semimetals. The reason for that is that in presence of the gap nodes the dynamics is dissipative, and cannot be described using a Lagrangian, which is only valid for conservative, \jaakko{non-dissipative} dynamics.
This is the main difference from the 3D QHE in a  (weak topological)  insulator, where the AQHE is not dissipative, and the description in terms of the action is appropriate. The action in Eq.(\ref{action}) does not violate the gauge invariance.

In relativistic theories, the gauge invariance is not violated if  $q_\mu$  is a fundamental constant of Nature, which does not change in space and time. However, such constant 4-vector violates the Lorentz invariance of the quantum vacuum \cite{Carroll1990}, as well as CPT symmetry.

\section{Conclusion}

The weak topological quantum Hall insulators in 3+1-dimensions are described by the anomalous quantum Hall effect and Chern-Simons term in the action in Eq. \eqref{action}. The AQHE response contains three elasticity vielbein fields $E^{~a}_\mu(x)$, with $a=1,2,3$.  In the absence of dislocations, the Chern-Simons action is gauge invariant.  In the presence of dislocations, the anomaly cancellation is produced by the Callan-Harvey mechanism \cite{RanEtAl09}.

The vielbeins $E^{~a}_\mu(x)$ entering the response have the same dimension as the vector-potential of the gauge field, i.e. the dimension $[L]^ {-1}$. This is consistent with the fact, that  in the presence of dislocations, the three  vielbein fields become the torsion gauge fields of the local $U(1) \times U(1) \times U(1)$ group of broken translation symmetry of crystals modulo the periodicity of the Bravais lattice \cite{BeekmanEtAl17}. The field strengths of these   gauge fields correspond to the torsion, expressed via tetrads according to Eq.(\ref{Torsion}). In general relativity, such description of vielbein in terms of the Lie group can be found in Ref. \cite{LL2}. Gravity based on tetrad fields with dimension  $[L]^ {-1}$ may occur also in the case that the microscopic structure of our quantum vacuum is periodic. However, at the moment  it is clear that the Lorentz invariance persists at energy scale larger 
than the Planck scale, for linear perturbations in the photon dispersion relation \cite{Lorentz2017}.

In connection between the anomalies in spacetime dimensions $D=2n+2$, $D=2n+1$ and $D=2n$, the weak topological insulator system under consideration corresponds to $n=2$. In the presence of dislocations, the system can be described by the  mixed 5D Wess-Zumino term in Eq.(\ref{MixedAnomaly5D}), which in turn can be obtained by dimensional reduction from the 6D mixed anomaly term  in Eq.(\ref{MixedAnomaly6D}).  The response is sensitive to the topology in momentum-frequency space and the prefactor of the Chern-Simons and Wess-Zumino actions contains three integer topological invariants. These are the  winding number  3-forms  expressed in terms of the Green's function in momentum-frequency space.   We also noted that for Floquet systems, a 4D periodic description with four invariants becomes possible.
 
 Concerning the Weyl semimetals and Weyl superfluids and superconductors, we pointed out that for a general $q_{\mu}(x)$, the action in Eq. \eqref{actionWrong}
 is not gauge invariant.  In the presence of deformations of the parameter $q_{\mu}(x)$, an action corresponding to Eq. \eqref{actionWrong} would violate gauge invariance. It is only gauge invariant if $q_{\mu}$ is a constant of Nature in the corresponding topological vacuum. In this case, the Fermi arcs corresponding to the Weyl nodes can be seen by anomaly inflow in the presence of a domain wall in $q_{\mu}(x)$ \cite{RamamurthyHughes2015}. However, the Fermi arcs have a more robust derivation under deformations of the Fermi-surface and in terms of momentum conserving boundary conditions for the Weyl fermions in the semimetal \cite{Haldane,Witten15}. 
 
This is in contrast  to the chiral anomaly of Weyl fermions that has been experimentally observed in superfluid $^3$He-A.  Distinct from the 3+1d weak topological insulators with topological AQHE, the chiral anomaly of Weyl fermions cannot be represented in terms of a mixed 4D Chern-Simons action but arises from a 5D Wess-Zumino term of the effective gauge field of the orbital degrees of freedom with a integer quantization of the coefficient of the resultant topological term.

In the case considered here, the action  Eq. \eqref{action} contains the elasticity tetrads, which makes the action gauge invariant. The same action, if it is expressed in terms of the Weyl  fermion tetrads, looses  the gauge invariance. This demonstrates the principal difference between the elasticity tetrads and the Weyl  fermion tetrads.

Finally, as discussed in Sec. II, the effective gravity in terms of the Weyl tetrads emerging near the left handed and right handed chiral fermions in condensed matter  prompts the following question  for the interplay of chiral fermions of Standard Model and gravity:  It is possible that both gravity and the Weyl fermions of Standard Model are the fundamental phenomena. However, it is not excluded that both of them are emergent, or one of them is more fundamental than the other.

In the latter case  {there are}  two scenarios, depending on what is more fundamental: gravity  or the chiral Weyl fermions. If gravity is  more fundamental, the tetrads should be the same for left and right fermions. This means that  the Pauli matrices $\sigma^a$ with $a=1,2,3$ must have different signs for right and left  handed fermions. Conversely, if the chiral fermions are the primary objects, then the tetrads should be different for left and right handed fermions. But they must be connected by a (discrete) symmetry, in order to obtain the PT-invariant Dirac equation where the two components mix, and to produce the same metric field (standard unimetric gravity).

Here we did not consider the interplay of the Weyl tetrad gravity  and the elasticity tetrad gravity, which in principle may lead to new topological invariants, expressed for example via the combination of the two torsion fields. The effect of elastic deformations on the Weyl and Dirac fermions tetrads has been considered e.g. in
 Refs. \onlinecite{Vozmediano2007, Vozmediano2012, VolovikZubkov2014,CortijoZubkov2016}.

This work has been supported by the European Research Council (ERC) under the European Union's Horizon 2020 research and innovation programme (Grant Agreement No. 694248). 

\begin{appendix}

\section{ Normalization of topological terms}

In this manuscript, we deal with \jaakko{the effective response of} chiral fermions coupled to Abelian gauge fields. More formally, we have line bundle with a $U(1)$-gauge connection on the $d$-dimensional manifold $M^d$ which has a spin structure. Let $A_{\mu}$ be the corresponding local gauge field and $F_{\mu\nu} = \partial_{\mu}A_{\nu}-\partial_{\nu}A_{\mu}$ its field strength. In a local gauge  $A = A_{\mu} dx^{\mu}$ and $F=d A =\frac{1}{2}F_{\mu\nu} dx^{\mu} \wedge dx^{\nu}$ in differential form notation, where the exterior derivative is the operator $d = (dx^{\mu} \wedge)\partial_{\mu}$. \jaakko{The exterior derivative satisfies $d^2 = 0$ and a closed $n$-form $h$ satisfies $d h = 0$ and is exact if $h=df$ for a $n-1$ form $f$}.  

The first and second Chern class \jaakko{of the $U(1)$ gauge field} are given as
\begin{align}
\textrm{ch}_1(F) = \frac{1}{2\pi}\int_{M^2} F \in \mathbb{Z}, \quad \textrm{ch}_2(F) = \frac{1}{4 \pi^2} \int_{M^4}  F \wedge F  \in \mathbb{Z}.
\end{align}
If $M^4$ is a spin manifold, the integer $\textrm{ch}_2(F)$ is even. The Chern-Simons functional is
\begin{align}
S_{\rm CS}[A] = \frac{1}{4\pi} \int_{M^3} A dA = \frac{1}{4\pi} \int_{M^3} d^3 x~ \epsilon^{\mu\nu\lambda} A_{\mu} \partial_{\nu} A_{\lambda}.
\end{align}
The local form of the Chern-Simons functional is not well defined, since if the line bundle is topologically non-trivial, one cannot pick a gauge that is valid everywhere even on a manifold without a boundary. The invariant definition is that one-extends $M^3$ to a four-manifold $M^4$, so that $\partial M^4 = M^3$. Then by partial integration using $\epsilon^{\mu\nu\alpha \beta}\partial_{\mu}(A_{\nu} \partial_{\alpha} A_\beta) = \frac{1}{4}\epsilon^{\mu\nu \alpha \beta} F_{\mu \nu}F_{\alpha \beta}$,
\begin{align}
S_{\rm CS}[A] = \frac{1}{16 \pi} \int_{M^4}d^4 x~ \epsilon^{\mu\nu\alpha\beta} F_{\mu\nu}F_{\alpha\beta}
\end{align}
If $A'$ and $M'^4$ is another such extension, 
\begin{align}
S_{4}[F] = S_{\rm CS}[A]-S_{\rm CS}[A'] = 2\pi \frac{1}{32\pi^2} \int_{M^4} d^4x~ \epsilon^{\mu\nu\alpha\beta} F_{\mu\nu}F_{\alpha\beta} =  \pi \textrm{ch}_2(F).
\end{align}  
Since we are studying fermions on spin manifolds, this is consistent with the integer quantization of the coefficient multiplying the CS action. 
\end{appendix}

 \end{document}